\documentclass[10pt]{llncs}
\usepackage{bold-extra}
\usepackage{color}
\usepackage{url}
\usepackage{graphicx}
\usepackage{listings}
\usepackage{multirow}
\usepackage{xspace}
\usepackage{tikz}
\usetikzlibrary{arrows,shapes}
\usetikzlibrary{plothandlers}
\begin{document}

%{{{ Title + Author + Abstract

\title{On Synergy of Metal, Slicing,\\ and Symbolic Execution}
\author{Ji\v{r}\'\i~Slab\'y, Jan~Strej\v{c}ek, and Marek~Trt\'{\i}k}
\institute{Faculty of Informatics, Masaryk University\\
  Botanick\'a 68a, 60200 Brno, Czech Republic\\
  \email{\{slaby,strejcek,trtik\}@fi.muni.cz}}

\maketitle

\begin{abstract}
  We introduce a novel technique for finding real errors in programs. The
  technique is based on a synergy of three well-known methods:
  metacompilation, slicing, and symbolic execution. More precisely, we
  instrument a given program with a code that tracks runs of state machines
  representing various kinds of errors. Next we slice the program to reduce
  its size without affecting runs of state machines. And then we
  symbolically execute the sliced program. Depending on the kind of symbolic
  execution, the technique can be applied as a stand-alone bug finding
  technique, or to weed out some false positives from an output of another
  bug-finding tool. We provide several examples demonstrating the practical
  applicability of our technique.
\end{abstract}

%}}}

%{{{ Intro

\section{Introduction}
\label{sec:introduction}

The title of this paper refers to two popular bug-finding techniques:
metacompilation and symbolic execution. The two techniques use completely
different principles leading to different advantages and disadvantages.

\begin{figure}[t]
  \centering
    \tikzstyle{loc} = [circle,thick,draw,minimum size=8mm]
    \tikzstyle{err} = [circle,thick,draw,minimum size=8mm,fill=red!10,draw=red]
    \tikzstyle{pre} = [<-,shorten <=1pt,>=stealth',semithick]
    \tikzstyle{post} = [->,shorten <=1pt,>=stealth',semithick]
    \begin{tikzpicture}[node distance=20mm]
        \node [loc] (U) {U};
        \node (S) [left of=U,xshift=10mm] {}
          edge [post] node {} (U);
        \node [loc] (L) [right of=U,xshift=10mm] {L}
          edge [pre,bend left=20] node [label=below:\texttt{lock($x$)}] {} (U)
          edge [post,bend right=20] node [label=above:\texttt{unlock($x$)}] {} (U);
        \node [err] (DU) [below of=U] {DU}
          edge [pre,near start] node [label=left:\texttt{unlock($x$)}] {} (U);
        \node [err] (DL) [below of=L] {DL}
          edge [pre,near start] node [label=left:\texttt{lock($x$)}] {} (L);
        \node [err] (RL) [below of=L,xshift=13mm] {RL}
          edge [pre,bend right=10,near start] node [label=right:\texttt{return}] {} (L);
    \end{tikzpicture}
    \caption{State machine $SM(x)$ describing errors in manipulation with
      lock $x$. The nodes U and L refer to states \emph{unlocked} and
      \emph{locked}, respectively. The other three nodes refer to error
      states: DU to \emph{double unlock}, DL to \emph{double lock}, and RL
      to \emph{return in locked state}. The initial node is U.}
  \label{fig:sm}
\end{figure}
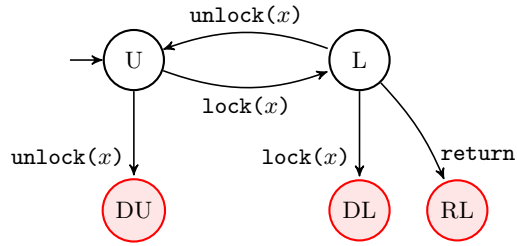

\emph{Metacompilation}~\cite{metal,pldi02} is a static analysis technique
looking for various kinds of errors specified by state machines. We explain
the technique with use of the state machine $SM(x)$ of Figure~\ref{fig:sm},
which describes errors in lock manipulation. Intuitively, the state machine
represents possible courses of states of a lock referenced by $x$ along an
execution of a program. The state of the lock is changed according to a
transition of the state machine if the execution performs a program statement
syntactically subsuming the label of the transition. We would like to decide
whether there exists any program execution where an instance of state machine
$SM(x)$ assigned to some lock of the analyzed program reaches an error state.
Unfortunately, this is not feasible due to potentially unbounded number of
executions and unbounded execution length. Hence, we use static analysis to
overapproximate the set of reachable states of state machines.

Let us assume that we want to check the program of Figure~\ref{fig:copy} for
errors specified by the state machine $SM(x)$. First, we find all locks in
the program and to each lock we assign an instance of the state machine. In
our case, there is only one lock pointed by \texttt{L} and thus only one
instance $SM(\texttt{L})$. For each program location, we compute a set
overapproximating possible states of $SM(\texttt{L})$ after executions leading
to the location. Roughly speaking, we initialize the set in the initial
location to $\{U\}$ and the other sets to $\emptyset$. Then we repeatedly
update the sets according to the effect of individual program statements until
the fixed point is reached.
% \item For each nonterminal program location $l$ and each state $s$ of set $S_l$,
%   if the program statement associated to the location syntactically subsumes
%   a label\footnote{Note that all occurence of \texttt{x} in the label are
%     replaced by the lock corresponding to the instance of the state
%     machine.} of some transition of $SM(\texttt{L})$ leading from $s$, then
%   we the target state of the transition has to be in the sets associated to
%   program locations reachable by the statement. Otherwise, the sets
%   associated to program locations reachable by the statement have to contain
%   the state directly the state $s$.
The resulting sets for the program of Figure~\ref{fig:copy} are written
directly in the code listing as comments.

\begin{figure}[!b]
  \centering
  \begin{tabular}{c}
    \begin{lstlisting}[language=C]
 1: char *copy(char *dst, char *src, int n, int *L) {
 2:    int i, len;                             // {U}
 3:    len = 0;                                // {U}
 4:    if (src != NULL && dst != NULL) {       // {U}
 5:       len = n;                             // {U}
 6:       lock(L);                             // {L}(*@\label{line:copy_lock}@*)
 7:    }                                       // {U,L}
 8:    i = 0;                                  // {U,L}
 9:    while (i < len) {                       // {U,L}
10:       dst[i] = src[i];                     // {U,L}
11:       i++;                                 // {U,L}
12:    }                                       // {U,L}
13:    if (len > 0) {                          // {U,L}
14:       unlock(L);                           // {DU,U}(*@\label{line:copy_unlock}@*)
15:    }                                       // {U,L}
16:    return dst;                             // {U,RL}(*@\label{line:copy_ret}@*)
17: }
    \end{lstlisting}
  \end{tabular}
    \caption{Function \texttt{copy} copying a source string \texttt{src}
      into a buffer \texttt{dst} using a lock \texttt{L} to prevent parallel
      writes.}
    \label{fig:copy}
\end{figure}

As we can see, the sets contain two error states: \emph{double unlock} after
the \texttt{unlock(L)} statement and \emph{return in locked state} in the
terminal location. If we analyze the computation of the sets, we can see
that the first error corresponds to executions going through lines
1,2,3,4,8, then iterating the \texttt{while}-loop and finally passing lines
13,14. These execution paths are not feasible due to the value of
\texttt{len}, which is set to 0 at line 3 and assumed to satisfy \texttt{len
  > 0} at line 13. Hence, the first error is a \emph{false positive}. The
second error corresponds to executions passing lines 1,2,3,4,5,6,7,8, then
iterating the \texttt{while}-loop and finally going through lines 13,16. All
these paths are also infeasible except the one that performs zero iterations
of the \texttt{while}-loop, which is the only real execution leading to
the only real locking error in the program.

To sum up, metacompilation is highly flexible, fast, and thus applicable on
extremely large software projects (e.g.~the Linux kernel). It examines all the
code and finds many error reports. Unfortunately, some of the reports are
false positives. The main source of false positives is related to the fact
that the analysis does not work with data values. In particular, the analysis
does not track connections between variable values and states of state
machines.
% Therefore, when a branching statement is processed, the analysis cannot
% decide which states of state machines are reachable on the positive branch
% and which on the negative one.
A drawback of this approach may be illustrated by the \emph{double unlock}
false positive detected in the program of Figure~\ref{fig:copy} because the
analysis does not know that the condition at line 13 holds only if the state
machine $SM(\texttt{L})$ is in state $L$.

At this point, we would like to emphasize that metacompilation actually uses
a more sophisticated algorithm enriched with many techniques for partial
elimination of false positives (see~\cite{pldi02} for
details). Metacompilation employs a dedicated language for description of
state machines called \textsc{Metal}.  The idea of error specification using
state machines appears in several tools including the original implementation
of metacompilation called \textsc{xgcc}~\cite{pldi02},
\textsc{Esp}~\cite{DLS02} or \textsc{Stanse}~\cite{stanse}.

\medskip In contrast to metacompilation, \emph{symbolic execution}~\cite{se} analyzes each execution path separately. In contrast to
standard execution, symbolic execution replaces input data by symbols
representing arbitrary values. Executed statements then manipulate
expressions over the symbols rather than exact values. For each execution
path, symbolic execution builds a formula called \emph{path condition},
which is a necessary and sufficient condition on input data to drive the
execution along the path. Whenever a path condition becomes unsatifiable,
the symbolic execution of this path is aborted as the path is unfeasible.
% Whenever symbolic execution reaches a branching
% statement, it uses the current path condition to check if any of the
% branches is unfeasible. If this is the case, the computation proceed to the
% other branch. Otherwise, the computation is forked and it continues in both
% branches. Moreover, we add the branching condition to the path condition on
% the positive branch and its negation to the path condition on the negative
% branch.
The main advantage of symbolic execution is that it works only with feasible
executions (assuming that we can decide satisfiability of a path condition)
and hence it does not report any false positives. A minor disadvantage is
that implementations of this technique usually detect only low-level errors
leading to crash. %(e.g.~segmentation fault).
To detect a specific kind of error, the program has to be modified to reduce
the error to a detected one (typically violation of an \texttt{assert}
statement). The main disadvantage of the technique is its high compuation
cost. In particular, programs containing loops or recursion have typically
large or even infinite number of execution paths and cannot be completely
analyzed by symbolic execution. Hence, symbolic execution usually explores
only a part of an analyzed program.

\medskip In this paper, we introduce a new technique offering a flexibility
of metacompilation and zero false positive rate of symbolic execution. The
basic idea is very simple: we use the concept of state machines to get
flexibility in error specification. Then we instrument a given program with
a code for tracking behaviors of the state machines. The instrumented
program is then reduced using the slicing method introduced
in~\cite{Wei84}. The sliced program has to meet the criterion to be
equivalent to the instrumented program with respect to reachability of error
states of tracked state machines. Note that slicing may remove big portions
of the code, including loops and function calls. Hence, an original program
with an infinite number of execution paths may be reduced to a program with
a finite number of execution paths. Finally, we execute the sliced program
symbolically.

Our technique may be used in two ways according to the applied symbolic
execution tool. If we apply a symbolic executor that prefers to explore more
parts of the code (for example, it can explore only the execution paths
iterating each program loop at most twice), we may use the technique as a
general bug-finding technique reporting only real errors. Note that this
approach may miss errors appearing only on unexplored paths. On the
contrary, if we use a symbolic executor exploring all execution paths, we
may use our technique for basic classification of error reports produced by
other tools (e.g.~\textsc{xgcc} or \textsc{Stanse}). For each such an error
report, we may instrument the corresponding code only with the state machine
describing that reported error. If our technique finds the same error, it is
a real one. If our technique explores all execution paths of the sliced code
without detecting the error, it is a false positive. If our technique runs
out of resources, we cannot decide whether the error is a real one or just a
false positive.

We have developed an experimental tool implementing our technique. The tool
instruments a program with a state machine describing locking errors (we use
a single-purpose instrumentation so far), then it applies an interprocedural
slicing to the instrumented code, and it passes the sliced code to symbolic
executor \textsc{Klee}~\cite{klee}. Our experimental results indicate that
the technique can indeed classify error reports produced by \textsc{Stanse}
applied to the Linux kernel.

\medskip We emphasize the synergy of the three known methods combined in the
presented technique.
\begin{itemize}
\item The errors are specified by state machines (inspired by
  metacompilation) and a given program instrumented with a code emulating the
  state machines. This provides us simple slicing criteria: we want to
  preserve values of memory places representing states of state machines.
  Hence, the sliced program contains only the code relevant to the considered
  errors.
\item Slicing may substantially reduce the size of the code, which in turn may
  remarkably improve performance of the symbolic execution.
\item Application of symbolic execution brings us another benefit. While in
  metacompilation, the state machines are associated to syntactic objects
  (e.g.~lock variables appearing in a program), we may associate state
  machines to actual values of these objects. This leads to a higher precision
  of error detection, which may potentially result in a detection of real
  errors missed by metacompilation. 
\end{itemize}

The rest of the paper is organized as follows.
Sections~\ref{sec:instr},~\ref{sec:slicing}, and~\ref{sec:symexe} deal with
program instrumentation, slicing, and symbolic execution, respectively.
% Section~\ref{sec:BackMapping} discusses extraction of
% counterexamples.\todo{Chceme to tu nechat?}
Experimental implementation of our technique and some experimental results
are discussed in Section~\ref{sec:impl}. Section~\ref{sec:related} is
devoted to related work while Section~\ref{sec:future} indicates some
directions driving our future research. Finally, the last section summarizes
presented results.

\section{Instrumentation} \label{sec:instr}

The purpose of the instrumentation phase of our algorithm is to insert a
code implementing a state machine into the analysed program. Nevertheless,
the semantics of the program being instrumented must not be changed.  A
result of this phase is therefore a new program that still has the original
functionality and it simultaneously updates instrumented state machines. We
show the process using the state machine $SM(x)$ of Figure~\ref{fig:sm} and
the program consisting of a function \texttt{foo} of~\ref{fig:foo} and the
function \texttt{copy} of Figure~\ref{fig:copy}. The function \texttt{foo}
calls the function \texttt{copy} twice, first with the lock \texttt{L1} and
then with the lock \texttt{L2}. The locks guard writes into buffers
\texttt{buf1} and \texttt{buf2} respectively. The function \texttt{foo} is
a so-called \emph{starting function}. It is a function where the symbolic
execution starts.

\begin{figure}[!htb]
  \centering
  \begin{tabular}{c}
    \begin{lstlisting}[language=C]
char *buf1, *buf2;
int L1, L2;

void foo(char *src, int n) {
   copy(src, buf1, n, &L1);
   copy(src, buf2, n, &L2);
}
    \end{lstlisting}
  \end{tabular}
  \caption{Function \texttt{foo} forms the analysed program together
    with function \texttt{copy}.}
  \label{fig:foo}
\end{figure}

The instrumentation starts by recognizing the code fragments in the analysed
program which manipulate with locks. More precisely, we look for all those
code fragments matching edge labels of the state machine $SM(x)$ of
Figure~\ref{fig:sm}. The analysed program contains three such fragments, all
of them in function \texttt{copy} (see Figure~\ref{fig:copy}): the call to
\texttt{lock} at line~\ref{line:copy_lock}, the call to \texttt{unlock} at
line~\ref{line:copy_unlock}, and the return statement at
line~\ref{line:copy_ret}.

Next we determine a set of all locks that are manipulated by the program.
From the recognized code fragments% (and according to the structure
% of edge labels of the state machine $SM(x)$ of Figure~\ref{fig:sm})
, we find out that a pointer variable \texttt{L} in \texttt{copy} is the
only program variable through which the program manipulates with locks.
% (We assume the code represented by dots in \texttt{foo} is not involved in
% the lock manipulation.)
Using a points-to analysis, we obtain obtain the set $\{ \texttt{L1},
\texttt{L2} \}$ of all possible locks the program manipulates with.
% Therefore, from \texttt{L}'s points-to set we obtain a set $\{ \texttt{L1},
% \texttt{L2} \}$ of all possible locks the program manipulates with.

We introduce a unique instance of the state machine $SM(x)$ for each lock in
the set. More precisely, we define two integer variables \texttt{smL1} and
\texttt{smL2} for keeping current state of state machines $SM(\texttt{L1})$
and $SM(\texttt{L2})$, respectively. Further, we need to specify a mapping
from locks to their state machines. The mapping is basically a function
(preferably with constant complexity) from addresses of program objects
(i.e.~the locks) to addresses of related state machines.
Figure~\ref{fig:fire} shows an implementation of a function
\texttt{smGetMachine} that maps addresses of locks \texttt{L1} and
\texttt{L2} to addresses of related state machines. We note that the
implementation of \texttt{smGetMachine} would be more complicated if state
machines are associated to dynamically allocated objects.

% Note that lines marked by \texttt{*} represent the code we have
% instrumented into the original program.

% \begin{figure}[!htb]
%   \centering
%   \begin{tabular}{c}
%     \begin{lstlisting}[language=C]
%  1: int smL1, smL2;
%  2:
%  3: int *smGetMachine(int *p) {
%  4:    if (p == &L1) return &smL1;
%  5:    if (p == &L2) return &smL2;
%  6:    return NULL;   // unreachable
%  7: }
%     \end{lstlisting}
%   \end{tabular}
% 	\caption{A mapping from locks to related state machines.}
% 	\label{fig:mapping}
% \end{figure}

\begin{figure}[p]
  \centering
  \begin{tabular}{c}
    \begin{lstlisting}[language=C]
 1: const int smU  = 0;     // state U
 2: const int smL  = 1;     // state L
 3: const int smDU = 2;     // state DU
 4: const int smDL = 3;     // state DL
 5: const int smRL = 4;     // state RL
 6:
 7: const int smLOCK   = 0; // transition lock(x)
 8: const int smUNLOCK = 1; // transition unlock(x)
 9: const int smRETURN = 2; // transition return
10:
11: int smL1 = smU, smL2 = smU;
12:
13: int *smGetMachine(int *p) {
14:    if (p == &L1) return &smL1;
15:    if (p == &L2) return &smL2;
16:    return NULL;         // unreachable
17: }
18:
19: void smFire(int *SM, int transition) {
20:    switch (*SM) {
21:    case smU:
22:       switch (transition) {
23:       case smLOCK:
24:          *SM = smL;
25:          break;
26:       case smUNLOCK:
27:          assert(false); // double unlock
28:          break;
29:       default: break;
30:       }
31:       break;
32:    case smL:
33:       switch (transition) {
34:       case smLOCK:
35:          assert(false); // double lock
36:          break;
37:       case smUNLOCK:
38:          *SM = smU;
39:          break;
40:       case smRETURN:
41:          assert(false); // return in locked
42:          break;
43:       default: break;
44:       }
45:       break;
46:    default: break;
47:    }
48: }
    \end{lstlisting}
  \end{tabular}
  \caption{Implementation of the state machine (\texttt{smFire}) and its
    identification (\texttt{smGetMachine}).}
  \label{fig:fire}
\end{figure}

% Although we have already instrumented the program with state machines, there
% is no code manipulating them.
Figure~\ref{fig:fire} contains also many constants and a function
\texttt{smFire} implementing the state machine $SM(x)$. Further,
Figure~\ref{fig:fire} declares variables \texttt{smL1} and \texttt{smL2} and
initialize them to the initial state of the state machine.  Note that we
represent both states of the machine and names of transitions by integer
constants. Also note that the pointer argument \texttt{SM} of
\texttt{smFire} function points to an instrumented state machine, whose
transition has to be fired.

It remains to instrument the recognized code fragments in the original
program. For each fragment we know its related transition of the state
machine and we also know what objects the fragment manipulates with (if
any). Therefore, we first retrieve an address of state machine related to
manipulated objects (if any) by using the function \texttt{smGetMachine} and
then we fire the transition by calling the function \texttt{smFire}. The
instrumented version of the original program consists of the code of
Figure~\ref{fig:fire} and the instrumented version of the original functions
\texttt{foo} and \texttt{copy} given in Figure~\ref{fig:copy_instrumented},
where the instrumented lines are highlighted by~\texttt{*}.

\begin{figure}[!htb]
    \begin{center}
    \begin{tabular}{l}
    \begin{lstlisting}[language=C]
  char *buf1, *buf2;
  int L1, L2;

  char *copy(char *dst, char *src, int n, int *L) {
     int i, len;
     len = 0;
     if (src != NULL && dst != NULL) {
        len = n;
*       smFire(smGetMachine(L), smLOCK);
        lock(L);
     }
     i = 0;
     while (i < len) {
        dst[i] = src[i];
        i++;
     }
     if (len > 0) {
*       smFire(smGetMachine(L), smUNLOCK);
        unlock(L);
     }
*    smFire(smGetMachine(L), smRETURN);
     return dst;
  }

  void foo(char *src, int n) {
     copy(src, buf1, n, &L1);
     copy(src, buf2, n, &L2);
  }
    \end{lstlisting}
    \end{tabular}
    \end{center}
    \caption{Functions \texttt{foo} and \texttt{copy} instrumented by calls
      of \texttt{smFire} function.}
    \label{fig:copy_instrumented}
\end{figure}

%\todol{doplnit recognition:} In practice we require a description language expressive enough to recognize wide range of possible expressions.
%\todol{doplnit constant function mapping (malloc):}

%}}}
%{{{ Slicing

\section{Slicing} \label{sec:slicing}

Let us have a look at the instrumented program in
Figure~\ref{fig:copy_instrumented}. We can easily observe, that the main
part of the function \texttt{copy}, i.e.~the loop copying the characters,
does not affect states of the instrumented state machines. Symbolic
execution of such a code is very expensive. Therefore, we use the slicing
technique~\cite{Wei84} to eliminate such a code from the instrumented
program.

The input of the slicing algorithm is a program to be sliced and a so-called
\emph{slicing criteria}. A slicing criterion is a pair of a program location
and a set of program variables. The slicing algorithm removes program
statements that do not affect any slicing criterion. More precisely, for
each input data passed to both original and sliced programs, values of the
variable set of each slicing criterion at the corresponding location are
always equal in both programs. Our analysis is interested only in states of
the instrumented automata, especially in locations corresponding to errors.
Hence, the slicing criterion is a pair of a location preceding an
\texttt{assert} statement in \texttt{smFire} function and the set of all
variables representing current states of the corresponding state machines. The
slicing criteria then consists of all such pairs.

\begin{figure}[!htb]
    \begin{center}
    \begin{tabular}{l}
    \begin{lstlisting}[language=C]
 1: char *buf1, *buf2;
 2: int L1, L2;
 3:
 4: char *copy(char *dst, char *src, int n, int *L) {
 5:    int len;
 6:    len = 0;
 7:    if (src != NULL && dst != NULL) {
 8:       len = n;
 9:       smFire(smGetMachine(L), smLOCK);
10:    }
11:    if (len > 0) {
12:       smFire(smGetMachine(L), smUNLOCK);
13:    }
14:    smFire(smGetMachine(L), smRETURN);
15:    return dst;
16: }
17:
18: void foo(char *src, int n) {
19:    copy(src, buf1, n, &L1);
20:    copy(src, buf2, n, &L2);
21: }
    \end{lstlisting}
    \end{tabular}
    \end{center}
    \caption{Functions \texttt{foo} and \texttt{copy} after slicing.}
    \label{fig:copy_sliced}
\end{figure}

In the instrumented program of Figures~\ref{fig:fire}
and~\ref{fig:copy_instrumented}, we want to preserve variables \texttt{smL1}
and \texttt{smL2}. We put slicing criteria into the lines of code detecting
transitions of state machines into error states. In other words, the slicing
criteria for our running example are pairs
(27,\{\texttt{smL1},\texttt{smL2}\}), (35,\{\texttt{smL1},\texttt{smL2}\}),
(41,\{\texttt{smL1},\texttt{smL2}\}), where the number refers to lines in the
code of Figure~\ref{fig:fire}. The result of the slicing procedure is
presented in Figures~\ref{fig:fire} and~\ref{fig:copy_sliced} (the code in
the former figure shall not changed by the slicing). Note that the sliced code
contains neither the \texttt{while}-loop nor the \texttt{lock} and
\texttt{unlock} commands.

It is important to note that some slicing techniques, including the one
in~\cite{Wei84} that we use, do not consider inputs for which the original
program does not halt.
% The problem arises when the sliced code contains an error, such that all
% executions exhibiting the error do not reach the error in the original
% code as they loop in some code that sliced.
Therefore, there is no way to guarantee that a sliced
program will fail to halt whenever the original program fails to halt. This
is the only principal source of potential false positives in our technique.

%}}}
%{{{ Symbolic Execution

\section{Symbolic Execution} \label{sec:symexe}

This is the final phase of our technique. We symbolically execute the sliced
program from the entry location of the starting function. Symbolic execution
explores real program paths. Therefore, if it reaches some of the assertions
inside function \texttt{smFire}, then we have found a bug.

Our running example nicely illustrates the crucial role of slicing to
feasibility of symbolic execution. Let us first consider symbolic execution
of the original program. It starts at the entry location of the function
\texttt{foo}. The execution eventually reaches the function
\texttt{copy}. Note that value of the parameter \texttt{n} is
symbolic. Therefore, symbolic execution will fork into two executions each
time we reach line~9 of Figure~\ref{fig:copy}. One of the executions skips
the loop at lines~9--12, while the other enters it. If we assume that the
type of \texttt{n} is a 32 bit integer, then the symbolic execution of one
call of \texttt{copy} explores more then $2^{31}$ real paths.

\begin{figure}[!tb]
    \centering
    \tikzstyle{loc} = [circle,thick,draw,inner sep=0.9pt,minimum size=4mm]
    \tikzstyle{loc2} = [circle,thick,draw,inner sep=0.1pt]%,minimum size=8mm]
    \tikzstyle{pre} = [<-,shorten <=1pt,>=stealth',semithick]
    \tikzstyle{post} = [->,shorten <=1pt,>=stealth',semithick]
    \footnotesize
    \begin{tikzpicture}[node distance=0.7cm]
        \node [loc] (0) {$19$};
        \node [loc] (1) [below of=0] {$6$}
            edge [pre] node {} (0);
        \node [loc] (2) [below of=1] {$7$}
            edge [pre] node {} (1);

        % ->false branch
        \node [loc] (3) [left of=2,below of=2,xshift=-0.5cm] {$11$}
            edge [pre] node [label=above:F~~] {} (2);
        \node [loc] (4) [below of=3] {$14$}
            edge [pre] node {} (3);
        \node [loc] (5) [below of=4] {$15$}
            edge [pre] node {} (4);
        \node [loc] (20) [below of=5] {$20$}
            edge [pre] node {} (5);
        \node [loc] (21) [below of=20] {$6$}
            edge [pre] node {} (20);
        \node [loc] (22) [below of=21] {$7$}
            edge [pre] node {} (21);
        \node [loc] (23) [right of=22,below of=22] {$8$}
            edge [pre] node [label=above:~~T] {} (22);
        \node [loc] (24) [below of=23] {$9$}
            edge [pre] node {} (23);
        \node [loc] (25) [below of=24] {$11$}
            edge [pre] node {} (24);
        \node [loc] (26) [left of=22,below of=22] {$11$}
            edge [pre] node [label=above:F~~] {} (22);
        \node [loc] (27) [below of=26] {$14$}
            edge [pre] node {} (26);
        \node [loc] (28) [below of=27] {$15$}
            edge [pre] node {} (27);
        \node [loc] (29) [left of=25,below of=25] {$14$}
            edge [pre] node [label=above:F~~] {} (25);
        \node [] (BUG) [below of=29,yshift=0.2cm] {\emph{bug}};
        \node [loc] (30) [right of=25,below of=25] {$12$}
            edge [pre] node [label=above:~~T] {} (25);
        \node [loc] (31) [below of=30] {$14$}
            edge [pre] node {} (30);
        \node [loc] (32) [below of=31] {$15$}
            edge [pre] node {} (31);

        % ->true branch
        \node [loc] (7) [right of=2,below of=2,xshift=0.5cm] {$8$}
            edge [pre] node [label=above:~~T] {} (2);
        \node [loc] (8) [below of=7] {$9$}
            edge [pre] node {} (7);
        \node [loc] (9) [below of=8] {$11$}
            edge [pre] node {} (8);

        % ->true->false branch
        \node [loc] (11) [left of=9,below of=9] {$14$}
            edge [pre] node [label=above:F~~] {} (9);
        \node [] (BUG) [below of=11,yshift=0.2cm] {\emph{bug}};

        % ->true->true branch
        \node [loc] (13) [right of=9,below of=9] {$12$}
            edge [pre] node [label=above:~~T] {} (9);
        \node [loc] (14) [below of=13] {$14$}
            edge [pre] node {} (13);
        \node [loc] (15) [below of=14] {$15$}
            edge [pre] node {} (14);

        \node [loc] (40) [below of=15] {$20$}
            edge [pre] node {} (15);
        \node [loc] (41) [below of=40] {$6$}
            edge [pre] node {} (40);
        \node [loc] (42) [below of=41] {$7$}
            edge [pre] node {} (41);
        \node [loc] (43) [right of=42,below of=42] {$8$}
            edge [pre] node [label=above:~~T] {} (42);
        \node [loc] (44) [below of=43] {$9$}
            edge [pre] node {} (43);
        \node [loc] (45) [below of=44] {$11$}
            edge [pre] node {} (44);
        \node [loc] (46) [right of=45,below of=45] {$12$}
            edge [pre] node [label=above:~~T] {} (45);
        \node [loc] (47) [below of=46] {$14$}
            edge [pre] node {} (46);
        \node [loc] (48) [below of=47] {$15$}
            edge [pre] node {} (47);
        \node [loc] (49) [left of=42,below of=42] {$14$}
            edge [pre] node [label=above:F~~] {} (42);
        \node [] (BUG) [below of=49,yshift=0.2cm] {\emph{bug}};

%        \node [] (X1) [above of=1,yshift=2mm] {}
%            edge [dotted,thick,draw] node {} (1);
    \end{tikzpicture}
    \caption{Symbolic execution tree of the sliced program of
      Figure~\ref{fig:copy_sliced}.}
        \label{fig:copy_sliced:SET}
\end{figure}
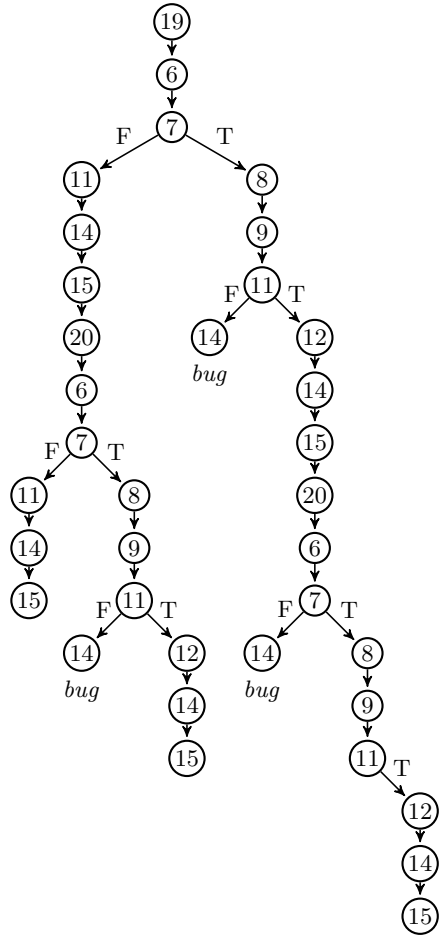

By contrast, the sliced program does not contain the loop, which generated
the huge number of real paths. Therefore, a number of real paths explored by
the symbolic execution is exactly 6. Figure~\ref{fig:copy_sliced:SET} shows
the symbolic execution tree of the sliced program of
Figure~\ref{fig:copy_sliced}. We left out vertices corresponding to lines in
called functions \texttt{smGetMachine} and \texttt{smFire}. Note that
although the parameter \texttt{n} has a symbolic value, it can only affect
the branching at line 11. Moreover, the parameter \texttt{L} always has a
concrete value. Therefore, we do not fork symbolic execution at branchings
inside functions \texttt{smGetMachine} and \texttt{smFire}. Three of the
explored paths are marked with the label \emph{bug}. These paths reach the
second assertion in function \texttt{smFire} (see Figure~\ref{fig:fire})
called from line 14 of the sliced program. In other words, the paths are
witnesses that we can leave the function \texttt{copy} in a locked
state. The remaining explored paths of Figure~\ref{fig:copy_sliced:SET} miss
the assertions in the function \texttt{smFire}. It means that the original
program contains only one locking error, namely \emph{return in locked
  state}.

% analysis.

%}}}
%{{{ Implementation

\section{Implementation and Experimental Results}
\label{sec:impl}
To verify applicability of the presented technique, we have developed an
experimental implementation. Our experimental tool works with programs in C
and, for the sake of simplicity, it detects only locking errors described by
a state machine very similar to $SM(x)$ of Figure~\ref{fig:sm}. The instances
of the state machine are associated with arguments of \texttt{lock} and
\texttt{unlock} function calls. Note that the technique currently works only
for the cases where a lock is instantiated only once during the run of the
symbolic executor. It works on a vast majority of the code we used. However we
plan to add a support even for the rest. The main part of our implementation
is written in three modules for the \textsc{Llvm} framework~\cite{llvm},
namely \texttt{Prepare}, \texttt{Slicer}, and \texttt{Kleerer}. The framework
provides us with the C compiler \textsc{\textsc{clang}}. We also use an
existing symbolic executor for \textsc{Llvm} called \textsc{Klee}~\cite{klee}.

Instrumentation of a given program proceeds in two steps. Using a C
preprocessor, the original program is instrumented with function calls
\texttt{smFire} located just above statements changing states of state
machines. The program is then translated by \textsc{clang} into \textsc{Llvm}
bytecode~\cite{llvm}. Optimizations are turned off as required by
\textsc{Klee}. The rest of the instrumentation (e.g. adding global variables
and changing the code to work with them) is done on the \textsc{Llvm} code
using the module \texttt{Prepare}.

The module \texttt{Slicer} implements a variant of the inter-procedural
slicing algorithm by Weiser~\cite{Wei84}. To guarantee correctness and to
improve performance of slicing, the algorithm employs points-to analysis by
Andersen~\cite{andersen}.

The module \texttt{Kleerer} performs a final processing of the sliced
bytecode before it is passed to \textsc{Klee}. In particular, the module
adds to the bytecode a function \texttt{main} that calls a \emph{starting
function}.  The \texttt{main} function also allocates a symbolic memory for
each parameter of the starting function. Size of the allocated memory is
determined by the parameter type. Plus, when the parameter is a pointer, the
size is multiplied by 4000. For example, 4 bytes are allocated for an integer
and 16000 bytes for an integer pointer. Further, for the pointer case, we pass
a pointer to the middle of the allocated memory (functions might dereference
memory at negative index). The idea behind is explained in~\cite{ucklee}.
Finally, the resulting bytecode is symbolically executed by~\textsc{Klee}. If
a symbolic execution touches a memory out of the allocated area, we get a
\emph{memory error}. To remedy this inconvenience, we plan to implement the
same on-demand memory handling \textsc{UcKlee}~\cite{ucklee} does.

%\begin{figure}[htb]
%\begin{center}
%\todol{Do we want this figure (picture needed)?} \\
%C $\rightarrow$ \textsc{Llvm} compiler (\textsc{clang}) $\rightarrow$
%Instrumentation (Prepare) $\rightarrow$ slicing (Slicer) $\rightarrow$
%main creator (Kleerer) $\rightarrow$ sym. exe. (Klee)
%\end{center}
%\caption{Phases of our tool}
%\label{fig:phases}
%\end{figure}

\subsection{Experiments}
\label{ssec:experiments}

\begin{table}[tb]
\begin{center}
\renewcommand*\arraystretch{1.2}
%\scriptsize
\begin{tabular}{|l|c|c|c|c|c|c|c|c|} \hline
  {\textbf{File}}&
  \multicolumn{5}{|c|}{\textbf{Running Time} (s)} & \textbf{Sliced} &
  \multirow{2}{*}{\textbf{Result}} & \textbf{Factual} \\ \cline{2-6}
  {\textbf{Function}} & \textbf{Comp.} & \textbf{Instr.} &
   \textbf{Slic.} & \textbf{SE} & \textbf{Total} & (\%) &
 & \textbf{State} \\ \hline \hline

  \begin{minipage}{3cm}
    \smallskip\texttt{\scriptsize fs/jfs/super.c}\\[-1pt]
    \texttt{jfs\_quota\_write}\smallskip
  \end{minipage}
  & 1.25 & 0.18 & 0.15 & 5.09 & 6.67 & 67.8 & BUG & BUG \\ \hline

  \begin{minipage}{3cm}
    \smallskip\texttt{\scriptsize drivers/net/qlge/qlge\_main.c}\\[-1pt]
    \texttt{qlge\_set\_mac\_address}\smallskip
  \end{minipage}
  & 2.70 & 0.72 & 26.75 & 13.28 & 43.45 & 66.5 & BUG & BUG \\ \hline

  \begin{minipage}{3cm}
    \smallskip\texttt{\scriptsize drivers/hid/hidraw.c}\\[-1pt]
    \texttt{hidraw\_read}\smallskip
  \end{minipage}
  & 1.06 & 0.18 & 0.14 & \multicolumn{2}{|c|}{Timeout} & 67.0 & TO & BUG \\ \hline

  \begin{minipage}{3cm}
    \smallskip\texttt{\scriptsize drivers/net/ns83820.c}\\[-1pt]
    \texttt{queue\_refill}\smallskip
  \end{minipage}
  & 1.76 & 0.29 & 1.72 & 0.62 & 4.39 & 72.9 & FP & FP \\[-2pt] \hline

  \begin{minipage}{3.6cm}
    \smallskip\texttt{\scriptsize drivers/usb/misc/}\\[-3pt]
    \hspace*{2ex}\texttt{\scriptsize sisusbvga/sisusb\_con.c}\\[-1pt]
    \texttt{sisusbcon\_set\_palette}\smallskip
  \end{minipage}
  & 1.50 & 0.24 & 0.27 & 17.19 & 19.20 & 76.0 & FP & FP \\\hline

  \begin{minipage}{3cm}
    \smallskip\texttt{\scriptsize fs/jffs2/nodemgmt.c}\\[-1pt]
    \texttt{jffs2\_reserve\_space}\smallskip
  \end{minipage}
   & 1.04 & 0.18 & 0.22 & \multicolumn{2}{|c|}{Timeout} & 46.8 & TO & FP \\ \hline
  \begin{minipage}{3cm}
    \smallskip\texttt{\scriptsize kernel/kprobes.c}\\[-1pt]
    \texttt{pre\_handler\_kretprobe}\smallskip
  \end{minipage}
   & 0.32 & 0.09 & 0.51 & 2.43 & 3.35 & 66.3 & ME & FP \\ \hline
\end{tabular}
\end{center}
\caption{Experimental results. The table presents running time of
  preprocessing and compilation (\textbf{Comp.}), instrumentation including
  points-to analysis (\textbf{Instr.}), slicing (\textbf{Slic.}), symbolic
  execution (\textbf{SE}), and the total running time. The column
  \textbf{Sliced} presents the ratio of instructions sliced away from the
  instrumented \textsc{Llvm} code.  The column \textbf{Result} specifies the
  result of our tool: BUG means that the tool found a real error, FP means
  that the analysis finished without error found (i.e.~the original error
  report is a false positive), TO that the symbolic execution did not finish in
  time and ME denotes an occurrence of memory error. The last column specifies
  the factual state of the error report.}
\label{tab:results}
\end{table}

% fs/jfs/super.c jfs_quota_write 369 / 119 (67.75% is gone)
% drivers/net/qlge/qlge_main.c qlge_set_mac_address 1333 / 447 (66.47% is gone)
% drivers/net/ns83820.c queue_refill 1212 / 329 (72.85% is gone)
% drivers/usb/misc/sisusbvga/sisusb_con.c sisusbcon_set_palette 2936 / 705 (75.99% is gone)
% fs/jffs2/nodemgmt.c jffs2_reserve_space (to po 5m40s) 677 / 360 (46.82% is gone)
% drivers/hid/hidraw.c hidraw_read (to po 10m) 666 / 220 (66.97% is gone)
% kernel/kprobes.c pre_handler_kretprobe (mem err) 202 / 68 (66.34% is gone)

% klee -max-stp-time=10 -max-time=300
We have performed our experiments on several functions of the Linux kernel
2.6.28, where the static analyzer \textsc{Stanse} reported some error. More
precisely, \textsc{Stanse} reported an error trace starting in these
functions. We consulted the errors with kernel developers to sort out which
are false positives and which are real errors. All the selected functions
(and all functions transitively called from them) contain no assembler (in
some cases, it has been replaced by an equivalent C code) and no external
function calls after slicing.

We ran our experimental tool on these functions. All tests were performed on
a machine with an Intel E6850 dual-core processor at 3\,GHz and 6\,GiB of
memory, running Linux. We specified \textsc{Klee} parameters to time out after
10 seconds spent in an SMT solver and after 300 seconds of an overall running
time. Increasing these times brings no real effect in our environment. We do
not pass optimize option for \textsc{Klee} because it causes \textsc{Klee} to
crash for most of the input.
% The optimization might
% further speed the analysis up. The crashes are yet to be investigated.

% At this early stage of the tool, the evaluation is not performed on the whole
% kernel. It is because it still does not scale that much. We plan to extend
% the tool to be capable of handling whole kernel. We limited our testing to a
% set of files. We aimed the technique at the functions where abstract
% interpretation tools reported a trace. Be it a real error or false positive.
% We consulted the files with upstream developers to sort out which are in each
% category.

Table~\ref{tab:results} presents results of our tool on selected functions.
The table shows compilation, instrumentation, slicing, symbolic execution, and
the overall running time. Further, the table presents the ratio of
instructions that were sliced away from the instrumented \textsc{Llvm} code.
The last two columns specify the results of our analysis and the real state
confirmed by kernel developers. The table clearly shows that the bottleneck of
our technique is the symbolic execution.

Although the results have no statistical significance, it is clear that the
technique can in principle classify error reports produced by other tools like
\textsc{Stanse}. If our technique reports an error, it is a real one. If it
finishes the analysis without any error detected, the original error report is
a false positive. The analysis may also not finish in a given time, which is
usually caused by loops in the sliced code. Finally, it may report a memory
error mentioned above.

% In most cases with our experiments, the time spent by compilation,
% instrumentation and slicing may be silently ignored with respect to the SE.
% In the kernel, there are few exception like \texttt{qlge\_set\_mac\_address}
% presented here.  Slicing in this case is a bit expensive due to circular
% inter-procedural dependencies of the slicing criteria. So that many states are
% transferred over calls until the fixed point is reached. In any case, the time
% of slicing was not more than a minute.

% Other than that, the vast majority of the analysis time is spent with the SE.
% In some cases, the time limit is hit. In both functions in
% Table~\ref{tab:results} it it because of unbounded loops which depend on
% locks.

% Table~\ref{tab:results} also sketches an effect of the slicing. It can be seen
% that it is effective as it allows, in average, 66 per cent of the code to be
% removed.

%}}}
%{{{ Related Work

\section{Related Work} \label{sec:related}

A description of program properties in \textsc{Metal} language and
meta-level compilation is discussed in~\cite{CEH02,metal,ECC00,pldi02}. The
technique presented in~\cite{metal} found a thousands of bugs in real system
code. It provides an easy description of properties to be checked for and a
fast analysis.  Nevertheless, it suffers from false positives. Since false
positive rate has huge impact on practical usability, an important part of
the technique are false positive suppression algorithms like killing
variables and expressions, synonyms, false path pruning, and others. Besides
the suppression algorithms, bug-reports from the tool are further ranked
according to their probability of being real errors. There are generic and
statistical ranking algorithms ordering bug-reports. An extension introduced
in~\cite{ECH01} provides an automatic inference of some temporal properties
based on statistical analysis of assumed programmer's beliefs. The
\textsc{ESP}~\cite{DLS02} technique uses a similar language to
\textsc{Metal} for properties description. It implements an interprocedural
dataflow algorithm based on~\cite{RHS95} for error detection and an abstract
simulation pruning algorithm for false positives
suppression. \textsc{Stanse}~\cite{stanse}, a static analysis tool also uses
state machines for description of checked program properties. The
description is based on parametrised abstract syntax trees. Although this
tool found hundreds of real bugs in the Linux kernel, it suffers from a high
false positive rate since its false positive suppression algorithms are
very limited.

Program analysis tools based on symbolic execution~\cite{se} mainly discover
low-level bugs like division by zero, illegal memory access, assertion
failure etc. These tools typically do not have problems with false
positives, but they have problems with scalability to large programs. There
has been developed a lot of techniques improving the scalability to programs
used in practice. Modern techniques are mostly hybrid. They usually combine
symbolic execution with concrete
one~\cite{PKS05,GLM08:active_props,GLM08:fuzzing,GLE09,SMA05,TdH08}. There
are also hybrid techniques combining symbolic execution with a complementary
static analysis~\cite{Beckmanetal08,GNRT10,Gulavanietal06,GMR09,NRTT09}.
Symbolic execution can be accelerated by a compositional approach based on
function summaries~\cite{AGT08,G07}. Another approach to effective symbolic
execution introduced in~\cite{BCE08,klee,Cadar08} is based on recording of
already seen behaviour and pruning its repetition. The followng techniques
focus on reaching a specific program location. Fitnex~\cite{XTHS09}, a
search strategy implemented in \textsc{Pex}~\cite{TdH08}, guides a path
exploration to a particular target location using fitness function. The
function measures how close an already discovered feasible path is to the
target. The LESE~\cite{SPmCS09} approach introduces symbolic variables for
the number of times each loop was executed. The symbolic variables are
linked with features of a known grammar generating inputs. Using these
links, the grammar can control the numbers of loop iterations performed on a
generated input. A technique presented in~\cite{GL11} analyses loops
on-the-fly, i.e.~during simultaneous concrete and symbolic executions of a
program for a concrete input. The loop analysis infers variables that are
modified by a constant value in each loop iteration. These variables are
used to build loop summaries expressed in a form of pre and post
conditions. An algorithm in~\cite{ST11} constructs a nontrivial necessary
condition on input values to drive the program execution to a given
location. A technique presented in~\cite{OT11} introduces a pair of counters
for two different paths around loop for each recurrent variable. Each
counter keeps an information about the number of iterations around one path
since the last iteration around the other one. Finally, there is an
orthogonal line of research which tries to improve the symbolic execution
for programs with some special types of inputs. Some techniques deal with
programs manipulating strings~\cite{BTV09,XGM08}, and some other techniques
reduce input space using a given input grammar~\cite{GKL08,SPmCS09}.

The interprocedural static slicing was introduced by Weiser~\cite{Wei84}. But
nowadays, there are many different approaches to program slicing. They are
surveyed by several authors~\cite{BG,L01,T95}. Applications of slicing include
program debugging, reverse engineering and regression testing~\cite{GHS92}.

%}}}
%{{{ Future Work

\section{Future Work}
\label{sec:future}
Our future work has basically three independent directions.

First, we plan to run our tool to classify all lock-related error reports
produced by \textsc{Stanse} on the Linux kernel. The results should provide
a better image of practical applicability of the technique. To get a
relevant data, we should solve some practical issues like a correct
detection of starting functions, automatic replacement of assembler,
treatment of external function calls, etc. We should also implement an
on-demand memory allocation to \textsc{Klee} as discussed in
Section~\ref{sec:impl} or use a different executor.

The second direction is to adopt or design some convenient way for
specification of arbitrary state machines. It may be a dedicated language
similar to \textsc{Metal}. Then we plan to implement an instrumentation
treating these state machines. In particular, the instrumentation should
correctly handle state machines associated with dynamically allocated
objects.

Finally, we would also like to examine performance of our technique as a
stand-alone error-detection tool. To this point, we have to use a symbolic
executor aiming for maximal code coverage. In particular, such an executor
has to suppress execution paths that differ from explored paths only in
number of loop iterations. Unfortunately, we do not know about any publicly
available symbolic executor of this kind. However, it seems that
\textsc{UcKlee}~\cite{ucklee} (which is not public as of now) has been
designed for a similar purpose.

%}}}
%{{{ Conclusion

\section{Conclusion}
\label{sec:conclusion}

We have presented a novel technique combining three standard methods:
specification of errors with state machines, slicing, and symbolic execution.
We currently do not know about any technique combining arbitrary two of the
three methods. We have discussed a synergy of the three methods. Moreover,
our experimental results indicate that the technique can recognize some false
positives and some real errors in error reports produced by other
error-detection tools.

\paragraph{Acknowledgements}
\label{sec:acknowledgements}

Jan Strej\v{c}ek has been supported by The Czech Science Foundation
(GA\v{C}R), grants No.  P202/10/1469 and P202/12/G061. Ji\v{r}\'i Slab\'{y}
and Marek Trt\'{\i}k have been supported by The Czech Science Foundation
(GA\v{C}R), grant No.~102/09/H042.

%}}}

\bibliographystyle{plain}
\bibliography{sse}

\end{document}